# Relativistic Consistency Rules Out
# GRW-Pearle Collapse Schemes.

Casey Blood
Professor Emeritus of Physics, Rutgers University
CaseyBlood@gmail.com

## Abstract

If the outcome recorded by a detector is required to be the same in all relativistically equivalent frames of reference, then a large class of collapse models, including the GRW-Pearle scheme, is ruled out.

## Introduction

There can be several simultaneously existing versions of reality in the wave function (state vector) of quantum mechanics—Schrödinger's cat can be both alive and dead at the same time, for example. One proposed way to reconcile this with our perception of a single version of reality is to suppose the wave function collapses down to just one version. There is no experimental evidence for collapse [1-6]. However, it is still one of the major interpretations of quantum mechanics.

To see whether collapse is a viable interpretation from the theory side, we will examine a class of 'local' mathematical models of collapse which includes the Ghirardi-Rimini-Weber-Pearle (GRWP) scheme [7-13]. Our conclusion is that this class of models is in conflict with the reasonable principle that the outcome recorded by a detector must be the same in all relativistically equivalent frames of reference.

We first review a simplified form of the GRWP proposal. Then we use a split-beam photon experiment to show that in 'local' models of collapse, the outcome recorded by a detector will in general be different when viewed in frames of reference differing only by their velocities. The idea for using different relativistic frames of reference to show that collapse models give untenable results was suggested by Wechsler ([14] and private communication). But she applied the idea to single particle wave functions, where one does not expect collapse, rather than to the wave functions of detectors.

## The GRWP Model.

To illustrate the class of collapse models we are concerned with, we will review a simplified form of the mathematically elegant model of Pearle [7-12], which builds on the work of Ghirardi, Rimini, and Weber [13].

•Space is divided into a number of small cubes of approximate size $\alpha^3 = (10^{-5} cm)^3$. The number of particles (particle-like wave functions) in the $n^{\text{th}}$ volume is labeled $\eta_n$. For typical densities, this will be around $10^{10}$ if the volume is filled with matter, and it will be zero if the volume has no particles in it.

•There is a random variable $w_n(t)$ and a potential energy $i(\eta_n w_n - \lambda \eta_n^2)$ associated with each volume element, where $\lambda$ is a frequency on the order of $10^{-16}$ per second. Ignoring all the usual kinetic energy and interaction terms, this gives a Hamiltonian equation

$$\frac{i \partial \mid \Psi, t\rangle}{\partial t} = H_c \mid \Psi, t\rangle = i\left( \sum_n \left( \eta_n w_n(t) - \lambda \eta_n^2 \right) \right) \mid \Psi, t\rangle \qquad (1)$$

where the sum extends over all the volume elements in space.

•We work with a two-state wave function

$$\mid \Psi \rangle = a(1,t) \mid 1\rangle + a(2,t) \mid 2\rangle . \qquad (2)$$

The effect of the Hamiltonian of Eq. (1) is solely to change the values of the coefficients. The sum $\mid a(1,t)\mid^2 + \mid a(2,t)\mid^2$ does not stay constant in time (as it does in standard quantum mechanics) under this non-unitary Hamiltonian.

•The $w_n(t)$'s are chosen at random but with a bias towards making $\mid a(1,t)\mid^2 + \mid a(2,t)\mid^2$ larger. One can then show the mathematic implies that after $10^{-6}$ sec or so either $a(1,t)$ or $a(2,t)$ will become very large compared to the other, and this effectively introduces collapse (to the large-coefficient state), with probability $\mid a(i,0)\mid^2$ of collapse to state $i$.

•It is convenient to assume the detectors have a pointer which points either to *no* (no detection) or to *yes* (detection). If we add the detector states to Eq. (2), we have

$$\mid \Psi \rangle = a(1,t) \mid 1\rangle \mid D(1), yes\rangle \mid D(2), no\rangle +$$
$$a(2,t) \mid 2\rangle \mid D(1), no\rangle \mid D(2), yes\rangle \qquad (3)$$

If we consider detector D(1), we see that:
    • the $\eta$'s corresponding to the position of the pointer when the
    detector reads *yes* will have an approximate value of $10^{10}$ in the *yes*

state while the $\eta$'s corresponding to the position of the pointer when the detector reads *no* will have a value of 0;

• and the $\eta$'s corresponding to the position of the pointer when the detector reads *no* will have a value of 0 in the *yes* state and an approximate value of $10^{10}$ in the *no* state,

• with equivalent statements for detector D(2).

That is, the $\eta$'s have different values in the different versions of reality. This difference, along with the Hamiltonian of Eq. (1), is what drives the collapse.

## Experimental Setup.

A photon wave function is shot at a half-silvered mirror placed at $90^o$ to the beam. Part of the wave function continues through the mirror in the $+x$ direction and the rest is reflected and travels in the $-x$ direction, with wave function

$$a(1)\,|+x\rangle + a(2)\,|-x\rangle. \tag{4}$$

Detector D(1) is set up in the $+x$ direction and D(2) in the $-x$ direction, both at distance $d$ from the mirror.

**Relativity.** There are three 'events' in this experiment; event 0 when the beam is split, event 1, when part of the wave function is detected at D(1), and event 2, when the other part is detected at D(2). We will use three reference frames; reference frame 0 is attached to the mirror, reference frame A moves with velocity $v$ in the $+x$ direction and reference frame B moves with velocity $v$ in the $-x$ direction. The coordinates of event 0 are $t = 0$, $x = 0$ in all three frames. In frame 0, the coordinates of event 1 are $t = t_0 = d/c$, $x = d$, and the coordinates of event 2 are $t = t_0 = d/c$, $x = -d$.

Events 1 and 2 are simultaneous in frame 0 but not in frames A and B. To obtain the times in those frames, we use Lorentz transformations, with results

$$t_A(1) = \gamma\left(t_0 - vd/c^2\right)$$
$$t_A(2) = \gamma\left(t_0 + vd/c^2\right)$$
$$t_B(1) = \gamma\left(t_0 + vd/c^2\right) \tag{5}$$
$$t_B(2) = \gamma\left(t_0 - vd/c^2\right)$$
$$\gamma = \sqrt{1 - v^2/c^2}$$

So we see that events 1 and 2 occur at different times in the two frames. We suppose that $v$ is nearly equal to $c$ and that $d$ is large, so the reputed GRWP-like collapse is completed in a time less than $\delta t = 2\gamma v d / c^2$. (If the collapse time is about .1 ms and $v = .99c$, then $d$ must be about 1 km.)

## The Two Frames.

In frame A, detector D(1) is activated first ($t_A(1) < t_A(2)$). Thus the results of the GRWP collapse process will be determined solely by the dynamics at D(1) (because D(2) does not come into play during the collapse process in frame A). This implies the coefficients in frame A are determined solely by the $w$'s and $\eta$'s of detector D(1). That is,

$$a_A(1,t) = f_{A,1}\left(t, w_n^{D(1)}, \eta_n^{D(1)}\right),$$
$$a_A(2,t) = f_{A,2}\left(t, w_n^{D(1)}, \eta_n^{D(1)}\right) \tag{6}$$

Similarly the results of the GRWP collapse process in frame B will be determined solely by the dynamics at D(2). Thus

$$a_B(1,t) = f_{B,1}\left(t, w_n^{D(2)}, \eta_n^{D(2)}\right),$$
$$a_B(2,t) = f_{B,2}\left(t, w_n^{D(2)}, \eta_n^{D(2)}\right) \tag{7}$$

So in frame A, the $w^{D(1)}, \eta^{D(1)}$ determine which coefficient, $a(1)$ or $a(2)$, will win, whereas in frame B, the $w^{D(2)}, \eta^{D(2)}$, *which are completely independent of the* $w^{D(1)}, \eta^{D(1)}$ (because D(1) does not come into play in the collapse process in frame B), will determine the winner. And once a winner is chosen by the first detector, the mathematics implies it cannot be changed by the random processes in the second detector.

Thus it is entirely possible in this type of collapse that in frame A, $a(2)$ will effectively go to zero, which gives a reading of *yes* on D(1) and *no* on D(2) while in frame B, $a(1)$ could effectively go to zero, which gives a reading of *no* on D(1) and *yes* on D(2). That is, *there is no way to guarantee that the detectors as perceived in the two frames will yield the same result*. In fact, if $|a(1,0)|^2 = |a(2,0)|^2 = 1/2$, the detectors in the two frames will disagree on approximately half the runs (and this difference can be communicated between observers in the two frames).

This violates the eminently reasonable principle that the detectors in all relativistically equivalent frames should give non-contradictory results. Thus the GRWP collapse scheme is not acceptable.

## The Generality of the Result.

If we require that the outcome recorded by a detector must be the same in all relativistically equivalent frames, what does this imply for collapse schemes in general?  In models similar to GRWP, the mechanisms that cause collapse—the relevant specifications of the states (analogous to the $\eta_n$) and the random variables (analogous to the $w_n$) are *locally* attached to each detector.  Changing the $w$'s to fields [12,13] does not affect the argument because the fields and their effects are still local to a specific detector.  And I cannot see that Pearle's quasirelativistic quasilocal model [15] gets rid of the crucial locality.  To summarize:

•There can be no collapse unless a large number of particles are involved, for otherwise we risk losing the one-particle or few-particle interference patterns so typical of quantum mechanics.

•In the above experiment, it must be the macroscopic differences in the *yes* and *no* states of the detectors that are relevant to the collapse.

• In either frame, the collapse takes place at the first detector as if the second detector were not there.  Thus in frame A we have no reason to expect the random variables 'locally' associated with D(1) to be coordinated with the random variables locally associated with D(2) (with a similar statement for frame B).  This inevitably implies there will be contradictory results.  (As a further contradictory aspect of these models, note the mathematics implies there *will* be coordination of the $w$'s of the two detectors in frame 0.)

Therefore it does not seem possible to devise a collapse scheme in which there is rigorous coordination between the detector readings in frame A and frame B.  The only possibility seems to be some combination of collapse and hidden variables.  It might be that one branch of the *single-particle* (or few-particle) wave function is somehow singled out, as in the Bohm model [16,17], and there is then some process which causes the *multi-particle* (including the detectors) non-singled out branches of the wave function to collapse to zero.

We conclude that a collapse theory can be free of contradictions only if the information that determines the collapse is carried, Bohm-like, in the single-particle wave function.

## References

[1] Philip Pearle and Euan Squires, *Bound state excitation, nucleon decay experiments, and models of wave function collapse.* Phys. Rev. Lett., 73, (1993)

[2] 11. E. Garcia, *Results of a dark matter search with a germanium detector in the Canfran tunnel.* Phys. Rev. D**51**, 1458 (1995).

[3]. A. J. Leggett, *Testing the limits of quantum mechanics: motivation, state of play, prospects,* Journal of Physics: Condensed Matter **14** R415-R451 (2002).

[4] 9. O. Nairz, M. Arndt and A. Zeilinger, *Quantum interference experiments with large molecules.* Am. Jour. Phys. **71**, 319 (2003).

[5] 12. SNO collaboration, *Measurement of the total active $^8B$ solar neutrino flux at the Sudbury neutrino observatory with enhanced neutral sensitivity.* Phys. Rev. Lett. **92**, 181301 (2004).

[6] Philip Pearle, *How Stands Collapse II*, arXiv, quant-ph/0611212v3 (2007).

[7] P. Pearle, *Combining stochastic dynamical state-vector reduction with spontaneous localization.* Phys. Rev. A 39, 2277 (1989).

[8] G. C. Ghirardi, P. Pearle and A. Rimini, *Markov processes in Hilbert space and continuous spontanenous localization of systems of identical particles.* Phys. Rev. A 42, 78 (1990).

[9] Philip Pearle, *Ways to describe dynamical state-vector reduction.* Phys. Rev. A, 48, 913 (1993).

[10] P. Pearle, *Completely quantized collapse and consequences.* Phys. Rev. A 72 022112 (2005).

[11] Philip Pearle, *How Stands Collapse* I, arXiv, quant-ph/0611211v1 (2006).

[12] Philip Pearle.*Stress tensor for quantized random field and wave function collapse.* arXiv:quant-ph/0804.3427v2 (2008).

[13] G. C. Ghirardi, A. Rimini and T. Weber, *Unified dynamics for microscopic and macroscopic systems*, Phys. Rev. D**34**, 470 (1986); Phys. Rev. D**36**, 3287 (1987).

[14] Sofia Wechsler, *Does a measurement really collapse the wave function?*, arXiv:quant-ph/1002.4219 (2010).

[15] Philip Pearle, *Quasirelativistic quasilocal finite wave function collapse model,* arXiv:quant-ph/0502069 (2005).

[16] David Bohm, *A suggested interpretation of quantum theory in terms of "hidden" variables,* Phys. Rev. **85** 166,180 (1952).

[17] D. Bohm and B. J. Hiley *The Undivided Universe* (Routledge, New York, 1993).